\documentclass[11pt]{article}
\usepackage{amssymb}
\usepackage{epsfig}
\usepackage{sint}%,showlabels}
\usepackage{macros_static}

\begin{document}

\thispagestyle{empty}
\title{{\normalsize\vskip -50pt
\mbox{} \hfill DESY 07-140 \\
\mbox{} \hfill SFB/CPP-07-49 \\}
\vskip 25pt
Precision for B--meson matrix elements
}

\author{
\centerline{
            \epsfxsize=2.5 true cm
            \epsfbox{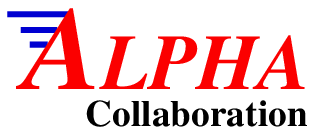}}\\
Damiano Guazzini$^{a}$, Rainer Sommer$^{a}$ and Nazario Tantalo $^{b}$
\\[1cm]
$^{a}$ DESY, Platanenallee 6, 15738 Zeuthen, Germany\\[0.5cm]
$^{b}$ INFN sezione "Tor Vergata", Via della Ricerca Scientifica 1, \\
  I-00133 Rome, Italy
        \&\\
        Centro E. Fermi, Compendio Viminale, I-00184 Rome, Italy
\\[1cm]
}

\maketitle

\begin{abstract}
  We demonstrate how HQET and the Step Scaling Method for B--physics, 
  pioneered by the Tor Vergata group, can be combined to reach a 
  further improved precision.
  The observables considered are
  the mass of the b-quark and the $\mrm{B_s}$--meson decay
  constant. The demonstration is carried out in quenched lattice QCD. 
  We start from a small volume, where
  one can use a standard O($a$)-improved relativistic action for the b-quark, 
  and compute two step scaling functions which relate the observables
  to the large volume ones. In all steps we extrapolate
  to the continuum limit, separately in HQET and in QCD for masses
  below $m_\mrm{b}$. The physical point $m_\mrm{b}$ is then reached by an interpolation
  of the continuum results in $1/m$. The essential, expected and verified, feature
  is that the step scaling fuctions have a weak mass-dependence resulting
  in an easy interpolation to the physical point.
  With $r_0=0.5\,\fm$ and the experimental $\mrm{B_\mrm{s}}$ and K masses
  as input, 
  we find $\fbs=191(6)$~MeV and the renormalization group invariant
  mass $M_\mrm{b}=6.88(10)$~GeV, translating into
  $\mbar_\mrm{b}(\mbar_\mrm{b})=4.42(6)$~GeV
  in the $\msbar$ scheme. This approach seems very promising for full QCD.

\end{abstract}

%\begin{flushright}
%  DESY 05-xxx \\
%  HU-EP-05/xx \\
%  SFB/CPP-05-xx
%\end{flushright}

\newpage

%%% Local Variables: 
%%% mode: latex
%%% TeX-master: "mbar"
%%% End: 

\section{Introduction}
\label{intro}

It has long been realized that B-meson decays
and mixing have a significant potential for the 
search for physics beyond the Standard Model of
particle physics. Unfortunately, the comparison of 
experimental results from BaBar and Belle to
the Standard Model has not yet revealed such effects.
An even higher precision in both future 
experiments and the corresponding ``predictions''
of the theory is required if we want to get hints
for new particles or interactions in this way.\footnote{In fact the 
situation on the theory side is not sufficiently 
clear to exclude that experiments have found new
physics already. The point is that hadronic matrix
elements of B-mesons are difficult to compute. It
could thus be that some matrix element (B-factor
or other) which has been extracted from fits
to the unitarity 
triangle is actually in disagreement with
the true matrix elements in QCD. Improved determinations
of these matrix elements are hence of interest even
without an increase of precision of the experiments.}

The most promising method for the computation of
QCD matrix elements with at most one hadron in 
initial and final states is lattice QCD. 
Contrary to what is sometimes reported,
it is, however, a very non-trivial task to achieve
precisions at the (few) percent level, keeping all
systematic uncertainties under control. This is 
particularly so in B-physics, where the difficulty of
simulating light quarks with masses that make contact to 
the regime where chiral perturbation theory is applicable
meets the additional requirement of correctly describing the physics 
of the heavy b-quark. The former requires lattices of
a large enough physical size, say $2-3\,\fm$ across and the 
latter a small lattice spacing, $a$, or the control 
of an effective theory (see 
\cite{lat03:kronfeld,reviews:pic03,reviews:hashimogi,lat06:onogi,nara:rainer} 
for more detailed accounts of the difficulties and recent
progress). In this letter we exclusively discuss
a method to cope with the discretization errors 
associated with the heavy quark dynamics. The light quark is simply
taken to be the strange quark and for the purpose of 
testing the methodology we work in the 
quenched approximation. The light (dynamical) quark
simulations are an entirely separate issue, where fortunately
significant progress has recently been made  
\cite{algo:GHMC,algo:L2,algo:stability,algo:urbach,algo:RHMC,algo:tautwo,algo:deflation,lat06:clark,algo:dwf}.

The basic idea of the approach investigated here 
is the fact that b-quarks
can be simulated (quite) straightforwardly in a space-time
volume with a linear extent of $L_0=\rmO(0.5\,\fm)$ 
\cite{lat01:rainer,lat02:rainer}. 
In such a volume the lattice
spacing can be chosen small enough such that
observables can be computed with a relativistic action
for the heavy quark. The continuum limit is reachable by
a short, controlled, extrapolation. Starting 
from this simple idea,
two different roads have been taken in the past 
\cite{hqet:pap1,Guagnelli:2002jd,hqet:pap4} and a 
third one has recently been explored \cite{RHQA:LC2}.

In the first the (continuum) observables in the small volume
serve to determine the parameters of HQET non-perturbatively
and then the physical
(large volume) matrix elements are computed in this effective theory.
By including $1/\mbeauty$-corrections a good overall precision is 
attainable
\cite{hqet:pap4}.

In the second way, one remains in the relativistic theory, and computes
the finite size effects of the observables iteratively 
($L_0\to L_1=sL_0\to L_2=s^2L_0\ldots$). As one
increases the volume also the lattice spacing is increased and
one has to reduce the mass, $m_\mrm{h}$,  of the actually simulated
quark to remain with $am_\mrm{h}\ll 1$. The physical mass of the 
b-quark is then reached by an extrapolation. 

Here we demonstrate how the two approaches can be combined by 
constraining the extrapolation to the physical quark mass with 
calculations in the effective theory; extrapolations are 
turned into interpolations and an even higher precision as well as
confidence is reached.

%%% Local Variables: 
%%% mode: latex
%%% TeX-master: "mbar"
%%% End: 

\section{Strategy}
\label{strat}

We are interested in computing an observable $O$,
which, in addition to the light quark masses, 
depends on the mass, $m_{\rm h}$,  of a heavy quark. 
Its exact definition will be mentioned when it becomes relevant.
In a Monte Carlo computation the observable depends in addition 
on the linear extent
$L$ of the simulated space-time volume. This finite size effect
is negligible when $L$ is large enough, 
which we here assume to be the case for $L \geq L_N$.  
Following \cite{romeII:mb,romeII:fb}, we express $O$
as a product of factors,
\be\label{eq:SSM_identity}
O(m_{\rm h},L_\infty)=O(m_{\rm h},L_0)\,
{ O(m_{\rm h},L_1) \over O(m_{\rm h},L_0)}\,
\cdots\, { O(m_{\rm h},L_N) \over O(m_{\rm h},L_{N-1})}\,.
\ee
Here $L_0$ is chosen small enough such that with an affordable
effort lattices with a spacing $a\ll 1/\mh$ can be used and
the continuum limit can be reached by an extrapolation
of $O$ computed with a relativistic $\rmO(a)$-improved action.
For the b-quark this means that  $a\approx0.012$~fm can be used.
In a small volume the details of the topology,
boundary conditions and the exact choice of observables are relevant.
We here note only that choosing \SF boundary conditions
makes such numerical computations affordable also when dynamical
quarks are included \cite{algo:GHMCalpha}. We come back to these details 
in \sect{obs}. 

The remaining factors in \eq{eq:SSM_identity} describe the
dependence on $L$. They are called step scaling functions. 
In their original version \cite{alpha:sigma}, they depended only 
on $L_i$ (or equivalently a renormalized coupling $\gbar(L_i)$),
but here we have an additional dependence on the mass of the 
heavy quark. It is convenient to replace the latter by the dimensionless 
{\it observable}
\be\label{eq:def_x}
x\equiv { 1 \over L\,\mps(m_{\rm h},L)}
= { 1 \over Lm_{\rm h}}
+ \rmO\left(\frac{ 1}{ \left(L\,m_{\rm h}\right)^2}\right)\,,
\ee
constructed from a finite volume pseudoscalar heavy-light mass, $\mps$.
It will then also be used as the HQET expansion parameter instead of the inverse 
of the heavy quark mass. 
The indicated HQET expansion % on the right-hand-side 
assumes that
$L$ is kept fixed; see e.g. \cite{hqet:pap1,nara:rainer} for more details; 
it will be used later.
First we define the generic step scaling function
\be\label{eq:sigma}
\sigma_{O}(x,L)\equiv
{ O(m_{\rm h},L) \over O(m_{\rm h},L/s)}\,, \quad
% x={ 1 \over L\, \mps(m_{\rm h},L)} \,,
\ee
(with $x$ from \eq{eq:def_x})
where the scale factor $s$ as well as any other quark masses
are kept fixed and are not indicated explicitly.

In particular, the step scaling function of the 
pseudoscalar mass itself, 
\be\label{eq:sigma_m}
\sigma_{\rm m}(x,L)\equiv
{ \mps(m_{\rm h},L) \over \mps(m_{\rm h},L/s)}\,, \quad
% x={ 1 \over L\,\mps(m_{\rm h},L)}\,,
\ee
is of central importance.
Starting from the experimentally determined mass,
$\mBs=5.3675(18)\,\GeV$ and $L_N$ large enough, it serves to
locate the physical points $x_i$ via 
\be\label{eq:x1_star}
  x_N = 1/(L_N\mBs)\,,\quad x_{i-1}=s\,\sigma_{\rm
    m}(x_i,L_i)\, 
  x_i\,.
\ee
The numerical results of all step scaling functions have to be
evaluated at these points. \Eq{eq:SSM_identity} is then rewritten 
as 
\be\label{eq:SSM}
O(m_{\rm h},L_\infty)=O(m_{\rm h},L_0)\, \sigma_O(x_1,L_1) \, 
\cdots\,  \sigma_O(x_N,L_N)\,.
\ee
Increasing $i$ in \eq{eq:x1_star}
successively, the computation of the 
step scaling functions in the relativistic theory {\em at the physical mass} 
requires lattice resolutions $L_i/a$ which become larger by a factor $s$
in each step. This is not affordable in practice. Thus the idea of 
\cite{romeII:mb,romeII:fb} was to compute $ \sigma_O(x,L_i)$ 
for a range of $x$ (and thus quark masses) such that $x\geq s^i x_i \approx x_0$
and to extrapolate $x\to x_i$. In other words in each step (starting from $L_0$)
the maximal quark mass which is simulated is reduced by about a factor $s$.
As expected from the fact that 
everywhere one is in the situation $x\ll1$,
the slopes in these extrapolations turned out to be rather small 
and the extrapolations could thus be carried out. % with some confidence.  
For an illustration of the $x$-dependence as it comes
out in practice, one may look ahead at our final results, \fig{f:interpol}.

Our main point in this paper is that these 
extrapolations can be 
turned into interpolations by computing
the limiting behavior for small $x$ directly in HQET,
\be\label{eq:SSM_expansion}
\sigma_{O}(x,L)=\sigma_{O}^{(0)}(L) +\rmO(x)
\,.
\ee
As usual in QCD, this large mass expansion is accompanied
by logarithms due to anomalous dimensions in the effective theory;
$\rmO(x)$ thus stands for at least one power of $x$ accompanied by
powers of $\log(x)$.
For $\sigma_{\rm m}$, the lowest order term 
is predicted by the theory to be one, while 
for the first order in $x$, a computation in the static
approximation of HQET ($\lag{stat}=\heavyb D_0 \heavy$) is required,
\be
 \label{eq:sigma_m_exp}
 \sigma_{\rm m}(x,L)= 1 + \sigma_{\rm m}^{\rm stat}(L)\,x +{\rm O}(x^2)\,.
\ee
The static term, which comes from the $\rmO(1/(Lm_{\rm h})^2)$ term in
\eq{eq:def_x}, is {\em not} accompanied by logarithms, see \sect{s:hqet}.
For the case of the decay constant already the lowest order
term is given by a non-perturbative computation in the static approximation,
\be
 \label{eq:sigma_f_exp}
 \sigma_{\rm f}^{(0)}(L) = \sigma_{\rm f}^{\rm stat}(L)\,.
\ee

As a further application of this method we compute the mass of the
b-quark starting from the physical meson mass $\mBs$.
To this end we define the ratio
\be\label{eq:rho}
\rho(x,L) \equiv
{ \mps(m_{\rm h},L) \over M_{\rm h}}% = \rho^{(0)}(L)+\rho^{(1)}(L)\, x+\Or(x^2)\,
\ee
of the meson mass to the renormalization group
invariant (RGI) quark mass, $M_{\rm h}$ (see e.g.~\cite{mbar:pap1}
for its definition). 
It provides  the connection 
\be\label{eq:b_mass}
M_{\rm b}={ \mBs \over \rho(x_0,L_0)\,
          \sigma_{\rm m}(x_1,L_1)\,\ldots\, 
          \sigma_{\rm m}(x_N,L_N)}\,.
\ee
between the physics input $\mBs$ and the RGI b-quark mass.

Note that the only
approximation made in the above equations is to neglect
finite size effects in the volume
of linear extent $L_N$.

\section{Finite volume observables}
\label{obs}
\subsection{Relativistic QCD \label{s:rel}}

Suitable finite volume observables are defined in the QCD  
\SF ~\cite{SF:LNWW,SF:stefan1}
with a space-time topology $L^3\times T$,
where $T=2L$ and $C=C'=0$ is chosen for the boundary gauge fields, and $\theta=0$ for
the phase in the spatial quark boundary conditions.

The \Oa-improved \cite{impr:SW,impr:pap1,impr:pap2,impr:pap3,impr:babp}
heavy-light correlation functions $\fa(t), \fp(t)$
and $f_1$ are defined and renormalized as in~\cite{romeII:mb}.
They are illustrated in \fig{f:correlfcts}.
%%%%%%%%%%%%%%%%%%%%%%%%%%%%%%%%%%%%%%%%%%%%%%%%%%%%%%%%%%%%
\FIGURE{
\includegraphics*[width=11.9cm]{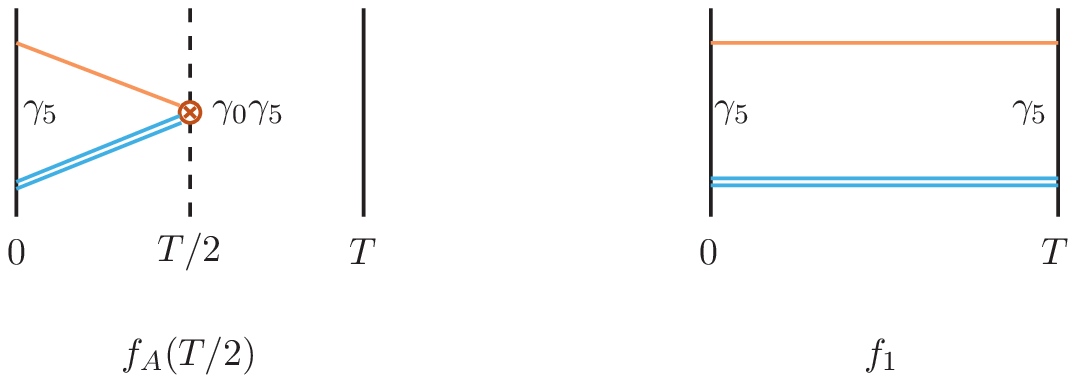}
\caption{\footnotesize
The boundary to axial-current correlator $\fa$ and the boundary to
boundary correlator $\fone$. Dirichlet boundary conditions are imposed at 
$x_0=0$ and $x_0=T$.
}\label{f:correlfcts}
}
%%%%%%%%%%%%%%%%%%%%%%%%%%%%%%%%%%%%%%%%%%%%%%%%%%%%%%%%%%%%%
They allow to define a finite volume pseudoscalar meson decay constant
and mass via 
\cite{mbar:pap2,impr:jochen}
\bes\label{eq:dc_qcd}
  \fps(m_{\rm h},L)&=&{ -2 \over \sqrt{L^3 \mps(m_{\rm h},L)}}
  {\fa(T/2) \over \sqrt{f_1}} \,,
\\
  \label{eq:meson_mass_qcd}
  \mps(m_{\rm h},L)&=&{1 \over 2a} [\ln(\fa(T/2-a))-\ln(\fa(T/2+a))]\,.
\ees
We remind the reader that we have a fixed ratio $T/L=2$. Therefore the
time separation in the correlation functions grows when $L$ grows.
Indeed, as discussed in detail in \cite{mbar:pap2}, these quantities
approach the physical ones in the large $L$ limit,
\bes
   \lim_{L\to\infty}  \mps(m_{\rm b},L) = \mBs\,,\qquad
   \lim_{L\to\infty}  \fps(m_{\rm b},L) = \fbs \,.
\ees
with corrections which (asymptotically) are exponentially small in $L$.
The associated step scaling functions are defined as
\bes
\label{eq:sigma_f}
\sigma_{\rm f}(x,L)={\fps(m_{\rm h},L)\sqrt{\mps(m_{\rm h},L)} \over \fps(m_{\rm h},L/s)\sqrt{\mps(m_{\rm h},L/s)}} \,,
\qquad x={1 \over L\,\mps(m_{\rm h},L)} \,,
\ees
and $\sigma_\mrm{m}$ as in \eq{eq:sigma_m}.

\subsection{HQET \label{s:hqet}}

In the static approximation of HQET,
unrenormalized correlation functions $\fastat$ and $\fonestat$
are defined in complete analogy to the relativistic ones
\cite{hqet:pap1} (see \cite{nara:rainer} for an introduction). 
As in these references,
we use the RGI
static axial current, related to the bare one
by a factor $\zastatrgi$. It serves to define the RGI ratio\,,
\be
\label{e:yrgi}
\YRGI(L)=\zastatrgi{\fastat(T/2) \over \sqrt{\fonestat(L)}}\,,
\ee
which is related to the QCD decay constant $\fps$ via
\be\label{eq:fb_hqet}
\fps(m_{\rm h},L)\sqrt{L^{3}m_{\rm PS}(m_{\rm h},L)}=
 -2\Cps(M_{\rm h}/\Lambda_\msbar)\times\YRGI(L)+\rmO(x)\,.
\ee
The function $\Cps(M_{\rm h}/\Lambda_\msbar)$, discussed in \cite{zastat:pap3,hqet:pap3},
originates from the matching of QCD and the effective theory.
In its numerical
evaluation we use the
anomalous dimension, $\gamma^{\rm PS}$ in the notation of \cite{hqet:pap3}. 
With the 3-loop term extracted from \cite{ChetGrozin} its uncertainty 
is estimated to be negligible~\cite{hqet:pap3} compared to our other errors.
Just like $\zastatrgi$,
it is needed only for $\fps(m_{\rm h},L_0)$;
it cancels out in the step scaling functions.

The pseudoscalar finite volume mass has an HQET expansion 
\cite{hqet:pap1}
\bes
  \label{eq:mps_hqet}
  m_{\rm PS}(m_{\rm h},L) = m_{\rm h} + \dmstat + \Gamma_{\rm stat}(L)
  + \rmO(1/m_{\rm h})\,,
\ees
with
\be
   \Gamma_{\rm stat}(L)={1 \over 2a} [\ln(\fastat(T/2-a))-\ln(\fastat(T/2+a))]
   \,,
\ee
where again we do not need to specify the renormalization scheme for
$m_{\rm h}$, but it is important that the counterterm $\dmstat$ cancels
the linear divergence in $\Gamma_{\rm stat}$ and the combination
$\dmstat + \Gamma_{\rm stat}(L)$ is of order $\Lambda_\mrm{QCD}$. Inserting
eqs.~(\ref{eq:fb_hqet},\ref{eq:mps_hqet}) into 
eqs.~(\ref{eq:sigma_f},\ref{eq:sigma_m}), we arrive at 
eqs.~(\ref{eq:sigma_f_exp},\ref{eq:sigma_m_exp}) with the 
static step scaling functions 
\bes\label{eq:stat_sigf}
\sigma_{\rm f}^{\rm stat}(L)&=&{1 \over s^{3/2}}
{\YRGI(L) \over \YRGI(L/s)}\,,
\\
\label{eq:stat_sigm}
\sigma_{\rm m}^{\rm stat}(L)&=&L\,[\Gamma_{\rm stat}(L)-\Gamma_{\rm stat}(L/s)]\,.
\ees
Here the renormalizations $\zastatrgi\times\Cps$ and $\dmstat$ cancel,
which shows that these static step scaling functions are {\em not}
accompanied by any logarithmic terms (in $m_\mrm{h}$ or $x$).
 
In our numerical investigation we will compute them precisely 
by using the static action denoted by HYP2 in \cite{stat:actpaper}
(see also \cite{HYP}),
and the corresponding $\Oa$-improvement 
coefficients for the static axial current.

\section{Results in the quenched approximation}
\label{res}

We employ the non-perturbatively \Oa-improved Wilson 
action \cite{impr:pap1,impr:pap3}.
The data at finite heavy quark mass are taken from
\cite{romeII:mb,romeII:fb}.
As there, we choose $N=2$ steps, $s=2$ and 
$L_0=0.4\,\fm$. The length scale is set by $r_0=0.5\,\fm$
\cite{pot:r0} using the parametrizations of $r_0/a$ as a function
of the bare coupling $g_0$ from \cite{pot:intermed,pot:r0_romeII}.
The light quark mass is set to the strange quark mass by fixing
the RGI-mass to $\Mstrange=0.1346(55)\,\GeV$ as previously
determined from the Kaon mass in the quenched approximation~\cite{mbar:pap3}.
The RGI-mass is related to the bare one by a non-perturbatively
computed renormalization factor $\zm$~\cite{mbar:pap1}, 
see e.g. \cite{romeII:mb} for details.

\begin{table}[h] 
{\small
 \centering
  \begin{tabular}{lclc}
   \hline\\[-1.0ex]
   $L[\fm]$ & $x$ & $\sigma_\mrm{m}(x,L)$ & $\sigma_\mrm{f}(x,L)$ 
\\[1.0ex]
   \hline\\[-1.0ex]
   1.6 & 0.0581&1.069(5) & 0.929(32)\\
       & 0.0670&1.081(6) & 0.912(27)\\
       & 0.0720 &1.087(7) & 0.900(24)
\\[1ex]
\hline\\[-1.0ex]
   0.8 & 0.0804&1.012(6) &0.4198(45)\\
       & 0.0884&1.014(6) &0.4193(45)\\
       & 0.1204&1.018(8) &0.4169(43)
\\[1ex]
\hline\\[-1.0ex]
   & & $\rho(x,L)$ & $\varphi(x,L)$
\\[1.0ex]
   \hline\\[-1.0ex]
   0.4 & 0.0933&0.744(09)  & 3.120(45)\\
       & 0.0990&0.754(09)  & 3.097(45)\\
       & 0.1472&0.837(12)  & 2.911(43)\\
       & 0.2768& - &2.534(40)\\
       & 0.2885& - & 2.505(40)
\\[1.0ex]
   \hline
  \end{tabular} 
 \caption{\small Finite mass observables after continuum extrapolation. 
          Physical units are set through $r_0=0.5\,\fm$. Statistical errors
          of $x$ due to $m_{\rm PS}$ have been changed to errors in
          the $x$-dependent observables.}
 \label{tab:ssf_rel}
}
\end{table}

\subsection{At finite heavy quark mass}
The data of \cite{romeII:mb,romeII:fb} have been reanalyzed.
The step scaling functions were first defined at a fixed
value of $r_0 M_\mrm{h}$ as in those references. Their continuum
limit was taken by an extrapolation linear in $(a/L)^2$, making use of 
different definitions of $M_\mrm{h}$ at finite lattice spacing and of
the fact that the continuum limit is independent of such details. 
Correlations between observables computed on the
same gauge configurations were taken into account.
The statistical uncertainties of the regularization dependent part of the
renormalization constants and the lattice spacing were included before
performing the continuum limit extrapolations, the uncertainty 
of the regularization independent 
part of the renormalization constants is added in the continuum limit;
all these do not appear as a
separate uncertainties, rather they are included in the quoted errors. 
For their detailed accumulation
we refer to \cite{thesis:damiano}. An impression on the
quality of the continuum extrapolations is easily obtained from the 
graphs in  \cite{romeII:mb,romeII:fb,thesis:damiano}. Since here our emphasis 
is on the use of the static approximation, we do not reproduce those details.
The continuum values of the step scaling functions were then interpolated
in the pseudoscalar mass to a few selected values of $x$. These are listed
in \tab{tab:ssf_rel} together with $\rho$, \eq{eq:rho} and
\bes
\label{e:phi}
 \varphi(x,L)=L^{3/2}F_{\rm PS}\sqrt{m_{\rm PS}}\,.
\ees

\begin{table}[p!] 
{\small 
 \centering
  \begin{tabular}{lllll}
   \hline\\[-1.0ex]
   $L_1/a$ & $\beta$ & $a\meffstat(L_1)$ & $a\meffstat(L_2)$ &
   $\sigma_\mrm{m}^{\rm stat}(L_2)$ % & $\sigma_\mrm{f}^\mrm{stat}(L)$ 
\\[1.0ex]
   \hline\\[-1.0ex]
    8 & 5.9598 & 0.3183(8) & & \\
    8.92& 6.0219 & 0.3000(5) & 0.4053(49) & 1.878(88) \\
   10 & 6.0914 & 0.2805(6) & & \\
   12 & 6.2110 & 0.2533(6) & & \\
   13.41& 6.2885 & 0.2360(5) & 0.3011(33) & 1.745(89) \\
   16 & 6.4200 & 0.2114(7) & & \\
   16.64& 6.4500 & 0.2067(6) & 0.2564(09) & 1.654(35) \\
   17.64& 6.4956 & 0.1997(6) & 0.2461(14) & 1.637(54) \\
   24 & 6.7370 & 0.1722(16) & & \\[1ex]
   \multicolumn{2}{c}{continuum}  & & & 1.561(53)
\\[1.0ex]
   \hline
  \end{tabular} 
 \caption{\small Numerical results in static approximation for
    $L_1=0.8\,\fm\,,\;L_2=2L_1$. The rows with non-integer $L_1/a$ list 
    interpolated values for $a\meffstat(L_1)$, while $a\meffstat(L_2)$
    are the large volume numbers of 
    \protect\cite{hqet:pap4,stat:letter,fb:interpol}.}
 \label{tab:static1}
}
\end{table}
\begin{table}[p!] 
{\small
 \centering
  \begin{tabular}{lllllllll}
   \hline\\[-1.0ex]
   ${L_0 / a}$ & $\beta$ & $a\meffstat(L_0)$ & $a\meffstat(L_1)$ &
   $\sigma_\mrm{m}^\mrm{stat}(L_1)$ & 
   $Y(L_0)$ & $Y(L_1)$ & $\sigma_\mrm{f}^\mrm{stat}(L_1)$ &  
\\[1.0ex]
   \hline\\[-1.0ex]
    6 & 6.2110  &  0.2272(9) & 0.2558(18) & 0.343(24) &
    -1.805(03) & -2.221(13) & 0.4350(27)\\
    8 & 6.4200  &  0.1958(9) & 0.2154(11) & 0.315(23) &
    -1.837(05) & -2.266(13) & 0.4361(26)\\
   12 & 6.7370 & 0.1561(8) & 0.1663(17) & 0.245(46) &
    -1.881(06) & -2.279(28) & 0.4284(55)\\
   16 & 6.9630 & 0.1355(7) & 0.1426(14) & 0.230(50) &
    -1.899(07) & -2.344(35) & 0.4366(67) \\
   24 & 7.3000 &  &  &  &
    -1.918(10) & &  \\[1ex]
   \multicolumn{2}{c}{continuum}   & & & 0.233(36) & & & 0.4337(44) 
\\[1.0ex]
   \hline
  \end{tabular} 
 \caption{\small Numerical results in static approximation for
    $L_0=0.4\,\fm$ and $L_1=2L_0$.}
 \label{tab:static2}
}
\end{table}
\begin{table}[p!] 
{\small
 \centering
  \begin{tabular}{lllllllll}
   \hline\\[-1.0ex]
   ${L_0 \over a}$ & $\beta$ & $\zastat({L_0 \over a},g_0)$ & 
   ${L_1 \over a}$ & $\beta$ & $\zastat({L_1 \over a},g_0)$
\\[1.0ex]
   \hline\\[-1.0ex]
    8 & 6.4200 & 0.8745(21) &12& 6.2110 & 0.7904(38)\\
   12 & 6.7370 & 0.8534(10) &16& 6.4200 & 0.7672(45)\\
   16 & 6.9630 & 0.8408(21) &24& 6.7370 & 0.7651(53)\\
   24 & 7.3000 & 0.8308(21) &32& 6.9630 & 0.7556(48)\\[1.0ex]
   \hline
  \end{tabular} 
 \caption{\small Renormalization factors for the static axial current at 
 renormalization scales $\mu=1/L_0$ and $\mu=1/L_1$ with
 $L_0=0.4\,\fm$ and $L_1=0.8\,\fm$.}
 \label{tab:static3}
}
\end{table}

\subsection{In static approximation}

We turn to the main new element in our numerical computations.
We start with the static step scaling function
$\sigma_\mrm{m}^\mrm{stat}(L_2)$,
requiring the computation of $\Gamma_{\rm stat}(L_1)$ and
$\Gamma_{\rm stat}(L_2)$, for several fixed values of $g_0$ followed by 
a continuum extrapolation. However, it is a central 
element of our strategy that $L_2 \approx 1.6\,\fm$ is  
large enough such that finite volume effects are negligible. Thus we can
replace $\Gamma_{\rm stat}(L_2)$ by $\Estat$, the ``mass'' of a static-strange 
bound state in large volume which is known from 
\cite{hqet:pap4,stat:letter,fb:interpol} in the 
range $6.0219\leq \beta=6/g_0^2 \leq 6.4956$. 
We have computed
$\Gamma_\mrm{stat}(L_1)$ for $L/a=8,10,12,16,24$, spanning
a wider range in $\beta$ and allowing easily for an interpolation to
the values of $\beta$ where $\Estat$ is known. All of this was done
for the HYP2 static action \cite{stat:actpaper} and 
for the tree-level as well as
the one-loop improved static-light axial current. Differences
between the two turned out to be far below our 
statistical precision of order $1-3\,\MeV$. The continuum extrapolation
of  $\sigma_\mrm{m}^\mrm{stat}(L_2)$, listed in \tab{tab:static1}, 
is well controlled, see \fig{f:contextrap}.

%%%%%%%%%%%%%%%%%%%%%%%%%%%%%%%%%%%%%%%%%%%%%%%%%%%%%%%%%%%%
\FIGURE{
\includegraphics*[width=14.5cm]{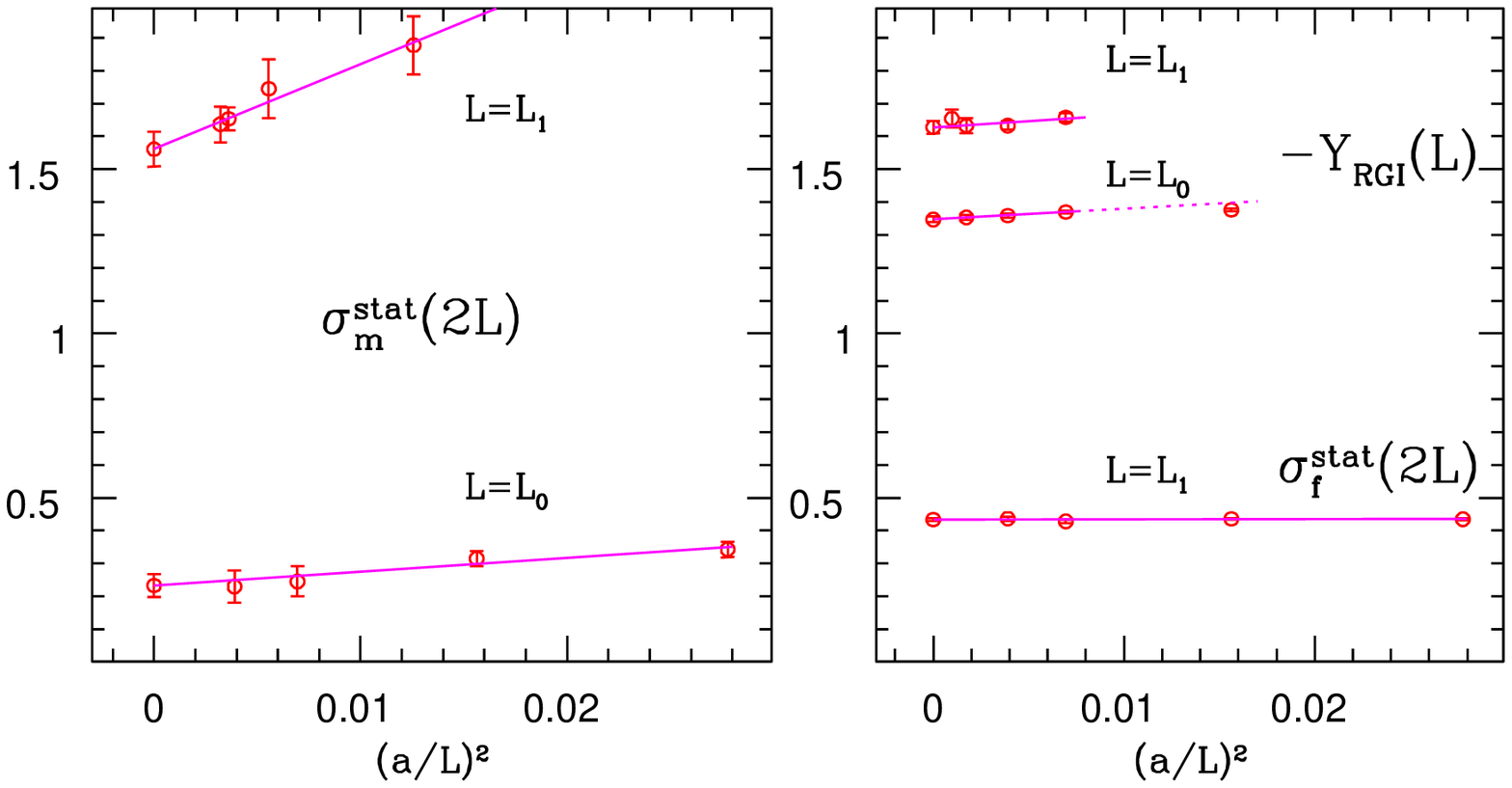}
\caption{\footnotesize
Continuum extrapolations of static results. The extrapolated values
with their errors are shown at $a/L=0$.
}\label{f:contextrap}
}
%%%%%%%%%%%%%%%%%%%%%%%%%%%%%%%%%%%%%%%%%%%%%%%%%%%%%%%%%%%%%

Similarly we profit from previous work in large volume
in the computation of $\sigma_\mrm{f}^\mrm{stat}(L_2)$, \eq{eq:stat_sigf}. 
In that case the continuum value 
\be\label{eq:dc_stat_Linfty}
\YRGI(L_2)=-4.65(19)\,,
\ee
is known from \cite{stat:letter,fb:interpol}. It remains
to compute
\bes
  \YRGI(L_1) &=& {\PhiRGI \over \PhiSF(\mu)} \times \zastat(L'/a,g_0) \times 
  Y(L_1/a,g_0)\,, \quad L'=1/\mu\,.
\ees
Here 
\bes
  Y(L/a,g_0)={\fastat(L) \over \sqrt{\fonestat(L)}}\,
\ees
is the unrenormalized version of \eq{e:yrgi} and
$\zastat(L'/a,g_0)$ is the factor, introduced
in \cite{zastat:pap3}, to renormalize the static axial
current in the (``new'') SF scheme, 
non-perturbatively at renormalization scale $\mu=1/L'$. Finally 
${\PhiRGI \over \PhiSF(\mu)}$ relates any matrix element of
the axial current in the SF scheme to the RGI matrix element. 
At the renormalization point $L'=L_1$ its non-perturbative value 
\be
  {\PhiRGI \over \PhiSF(1/L_1)} = 0.928(2) \,
\ee 
is easily extracted from the results in \cite{zastat:pap3}.
We have computed the missing factors $\zastat(L'/a,g_0)\,, Y(L_1/a,g_0)$
for various values of $L_1/a$, setting $L'=L_1$, see \tab{tab:static3}.
Note that following the exact definition of \cite{zastat:pap3}, 
$\theta=1/2\,, T=L'$ is employed for $\zastat$ and the computation is carried
out at zero (light) quark mass -- in contrast to the evaluation of $Y$ 
(and all other quantities).

The continuum limit (\fig{f:contextrap}) 
\be\label{eq:YRGI_2L0}
\YRGI(L_1)=-1.628(19)
\ee
is combined with (\ref{eq:dc_stat_Linfty}) to get
\be\label{eq:S2}
\sigma_{\rm f}^{\rm stat}(L_2)=1.010(43).
\ee

In the computation of the static step scaling functions 
$\sigma^\mrm{stat}(L_1)$ (\tab{tab:static2},~\fig{f:contextrap}) 
we followed straightforwardly their definitions.
Finally,  
\be
  {\PhiRGI \over \PhiSF(1/L_0)} = 0.846(6) \,.
\ee 
from \cite{zastat:pap3} together with 
$Y(L_0/a,g_0)\,,\zastat(L_0/a,g_0)$, Tables \ref{tab:static2} and 
\ref{tab:static3},
yields  
\be\label{eq:YRGI_L0}
\YRGI(L_0)=-1.347(13)\,
\ee
by a continuum extrapolation again illustrated in \fig{f:contextrap}.

\subsection{Interpolation to the physical point}

We now combine the static results with the relativistic ones,
through linear and quadratic interpolations in $x$. Namely we fit for
the parameters $m_{j}(L_i)$ and $e_j(L_i)$ in
\bes
   \label{e:mi}
   \sigma_m(x,L_i) &=& 1 + m_1(L_i)\,x + m_2(L_i)\,x^2\,,
   \\
   \label{e:m1}
   \sigma_{\rm m}^{\rm stat}(L_i) &=& m_1(L_i) \,,
   \\
   \label{e:ei}
   \sigma_{\rm f}(x,L_i) &=& e_0(L) + e_1(L_i)\,x + e_2(L_i)\,x^2\,,
\ees
and then insert the fit functions \eq{e:mi} and \eq{e:ei}
into \eq{eq:b_mass} and \eq{eq:fb_eq}. 
Note that the first two equations are fit together.  The static
$\sigma_{\rm m}^{\rm  stat}(L_i)$ enter \eq{e:m1} as data 
points and
$\sigma_{\rm f}^{\rm  stat}(L_i)$ are data at $x=0$ in \eq{e:ei}.
As seen in \fig{f:contextrap},
the quadratic terms are moderate in the whole range and
in particular at the physical points $x_i$ the differences 
between the static results and the interpolated 
ones are rather small. As an illustration of 
the effect of the static results we also carry out an analysis
 where they are {\em not} taken into account.
The numbers in \tab{tab:interpol} show that the statistical 
errors in the step scaling functions are significantly reduced by 
including the static constraints. Furthermore we can perform 
the consistency check of including quadratic terms only when the static 
constraints are used. The agreement between linear and quadratic 
interpolations is very reassuring.

\begin{table}[h] 
 \centering
  \begin{tabular}{lllllll}
   \hline\\[-1.0ex]
   $i$ & \multicolumn{2}{c}{$\sigma_\mrm{m}(x_i,L_i)$} & &
                 \multicolumn{2}{c}{$\sigma_\mrm{f}(x_i,L_i)$} &  Fit
\\[1.0ex]
   \hline\\[-1.0ex]
  2 &  \em 1.0330(11) & 1.0258(21) &&  0.985(31) &  & quadratic\\
  2 &  1.0319(11) & 1.0276(22) && \em 0.977(29) & 1.002(54) & linear\\[1ex]
  1 &  \em 1.0092(18) & 1.0074(33) && 0.4243(36) & & quadratic\\
  1 &  1.0093(15) & 1.0072(32) && \em 0.4260(31) & 0.4223(48) & linear 
\\[1.0ex]
   \hline
  \end{tabular} 
 \caption{\small Step scaling functions inter/extra-polated to the physical
          points $x_2=0.022974(8)$, $x_1=0.04746(5)$ and 
          $x_1^{\rm QCD}= 0.04741(10)$, where the latter originates from the 
          fits to only the finite heavy quark mass data. The 
          left--side column of each $\sigma$ is the number
          including the static constraint, the right--side one is
          without. }
 \label{tab:interpol}
\end{table}
%

%%%%%%%%%%%%%%%%%%%%%%%%%%%%%%%%%%%%%%%%%%%%%%%%%%%%%%%%%%%%
\FIGURE{
\includegraphics*[width=14.5cm]{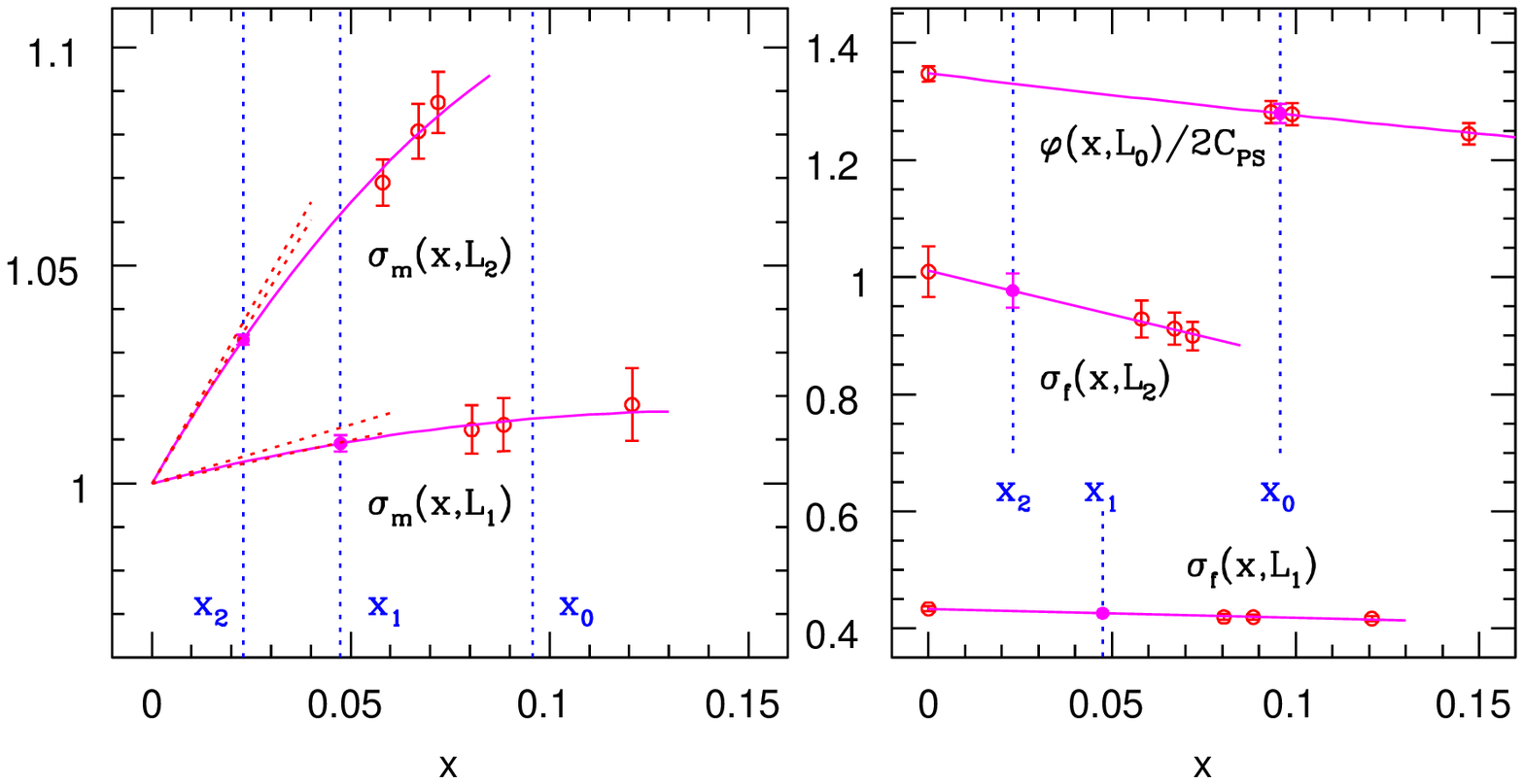}
\caption{\footnotesize
Interpolations to the physical points are shown 
by the filled circles. For $\sigma_\mrm{m}$,
the static constraints are illustrated as the error band
of the static result $1+\sigma_\mrm{m}^\mrm{stat} x$. On the right hand side,
the static results enter as data points at $x=0$.
}\label{f:interpol}
}
%%%%%%%%%%%%%%%%%%%%%%%%%%%%%%%%%%%%%%%%%%%%%%%%%%%%%%%%%%%%%

For $\rho(x,L_0)$ and $\fps(x,L_0)$ the relativistic simulations
straddle the physical point $x=x_0$ and, for the decay constant, 
the static data do not 
sensitively improve the precision on the interpolated point. 
However, as an
illustration how HQET does describe these quantities, we
also show \eq{eq:YRGI_L0} together with the data at finite $x$
in \fig{f:interpol}; in that case the interpolation yields  
$L_0^{3/2}\fps\sqrt{m_{\rm PS}}/(2\Cps)=1.279(17)$ or
$\varphi(x_0,L_0) = 3.107(41)$ and 
$\rho(x_0,L_0) =0.7485(9)$. 

Our final large volume results from \eq{eq:b_mass} and
\be\label{eq:fb_eq}
\fbs= \varphi(x_0,L_0)\,
          \sigma_{\rm f}(x_1,L_1)\, \sigma_{\rm f}(x_2,L_2)\,
          L_0^{-3/2}\, \mBs^{-1/2}
\ee
are
\be\label{e:res}
   \fbs=191(6)\,\MeV\,, \quad \Mb=6.88(10)\,\GeV\Longrightarrow
   \mbar_\mrm{b}(\mbar_\mrm{b})=4.42(6)\,\GeV \,.
\ee
Here the conversion to the running mass in the $\msbar$-scheme is done
with the 4-loop RG equations (for $\nf=0$ and
$\Lambda_\msbar^{(0)}=238(19)\,\MeV$ \cite{mbar:pap1}).

\section{Conclusions and outlook}
\label{concl}

We have followed a general strategy for computing 
B-meson observables. Starting from a finite volume,
where the observables are straightforwardly computable
in relativistic lattice QCD, we evaluated 
step scaling functions which describe the finite size effects.
The latter are not directly computable at the physical
points since for accessible lattices $a\mbeauty\geq1$. 
Previously these functions have either been computed by an 
extrapolation in the heavy quark mass to the physical 
$\mbeauty$ \cite{romeII:mb,romeII:fb} or they have been
computed in HQET \cite{hqet:pap1,hqet:pap4}. Here we have 
demonstrated how
the two approaches can be combined to further increase
precision and confidence in the results. 

\Fig{f:interpol}, which is a continuum graph, 
shows that the static (lowest order HQET) results
match very well onto the finite mass step scaling functions. 
We therefore have excellent control over the heavy
quark mass dependence
-- if desired from below the charm quark mass to 
the b-quark mass and beyond.

Our final numbers for decay constant and b-quark mass, \eq{e:res},
agree well with the previous estimates of 
\cite{romeII:mb,romeII:fb,hqet:pap4,stat:letter,fb:interpol}
where the same experimental data was used as input.\footnote{In 
\cite{hqet:pap4} the spin averaged $\mrm{B}_\strange$--mass was used instead
of the pseudoscalar mass, but this is a small effect of order 
$\Lambda_\mrm{QCD}^3/\mbeauty^2$.} 

In our results, \fig{f:interpol}, one notices that the corrections to the 
static approximation are very small at the b-quark mass. This represents
an intriguing demonstration of the precision and usefulness of HQET 
for B-physics. Although our exercise was in the quenched approximation,
such a qualitative result may well be carried over to (full) QCD.

Concerning the application of the strategy to QCD, 
the attentive reader will have noticed that 
in our computations we extensively relied on the knowledge of a
reference scale ($r_0/a$) over a large range of lattice spacings $a$.
This luxury is not available in full QCD -- and will
not be for a while to come. However, with the knowledge of the running coupling
of \cite{alpha:nf2}, one can properly set the scale also for 
small lattice spacings. We further note that the finite volume
computations which are needed in this strategy require a 
significantly smaller effort than the large volume ones. 

We therefore
conclude that the here investigated method is very promising for
the near future where we expect that high precision can be reached 
for B-physics. Note that the strategy may be extended to other
observables such as mass splittings \cite{Guazzini:2007bu,damiano:lat07} and form factors \cite{romeII:IW}.

\bigskip

{\bf \noindent Acknowledgement}

It is a pleasure to thank Michele Della Morte 
and Giulia Maria de Divitiis for useful discussions 
as well as some practical help. 
We thank NIC  for allocating computer time on the APEmille
computers at DESY Zeuthen to this project and the APE group 
for its help. This project has been supported by the DFG in the  
SFB Transregio 9 ``Computational Particle Physics'' and
by
the European community through
EU Contract No.~MRTN-CT-2006-035482, ``FLAVIAnet''.

% \newpage

\bibliography{refs}           %or whatever your .bib file is
\bibliographystyle{h-elsevier}   %if you use h-elsevier.bst

\end{document}